\documentclass[prl,twocolumn,amsmath,amssymb,showpacs,superscriptaddress,floatfix]{revtex4}

\usepackage{graphicx}
\usepackage{color}
\usepackage{soul}
\usepackage{epsfig}

\begin{document}
\title{Structures of local rearrangements in soft colloidal glasses}

\date{\today}

\author{Xiunan Yang}
\affiliation{Beijing National Laboratory for Condensed Matter Physics and Key Laboratory of Soft Matter Physics, Institute of Physics, Chinese Academy of Sciences, Beijing 100190, People's Republic of China}
\author{Rui Liu}
\affiliation{Beijing National Laboratory for Condensed Matter Physics and Key Laboratory of Soft Matter Physics, Institute of Physics, Chinese Academy of Sciences, Beijing 100190, People's Republic of China}
\author{Mingcheng Yang}
\affiliation{Beijing National Laboratory for Condensed Matter Physics and Key Laboratory of Soft Matter Physics, Institute of Physics, Chinese Academy of Sciences, Beijing 100190, People's Republic of China}
\author{Wei-Hua Wang$^*$}
\affiliation{Institute of Physics, Chinese Academy of Sciences, Beijing 100190, People's Republic of China}
\author{Ke Chen$^*$}
\affiliation{Beijing National Laboratory for Condensed Matter Physics and Key Laboratory of Soft Matter Physics, Institute of Physics, Chinese Academy of Sciences, Beijing 100190, People's Republic of China}

\begin{abstract}
We image local structural rearrangements in soft colloidal glasses under small periodic perturbations induced by thermal cycling. Local structural entropy $S_{2}$ positively correlates with observed rearrangements in colloidal glasses. The high $S_{2}$ values of the rearranging clusters in glasses indicate that fragile regions in glasses are structurally less correlated, similar to structural defects in crystalline solids. Slow-evolving high $S_{2}$ spots are capable of predicting local rearrangements long before the relaxations occur, while fluctuation-created high $S_{2}$ spots best correlate with local deformations right before the rearrangement events. Local free volumes are also found to correlate with particle rearrangements at extreme values, although the ability to identify relaxation sites is substantially lower than $S_{2}$. Our experiments provide an efficient structural identifier for the fragile regions in glasses, and highlight the important role of structural correlations in the physics of glasses.
\end{abstract}

\maketitle

Understanding the structure-property connections in glasses is one of the most challenging problems in condensed matter physics. The lack of a clear characterization of the structures in glasses has hindered the formulation of general theories for glass transition and deformation of glasses~\cite{stru1,stru2,stru3,stru4,stru5,stru6,stru7}. The most elementary relaxation events in glasses are the local atomic rearrangements that play similar roles in the mechanical properties of glasses as the structural defects in crystals. Local regions prone to rearrangements in glasses are often known as flow units~\cite{Wang1,Wang2}, shear transformation zones (STZs) \cite{Langer1,Argon1}, soft spots~\cite{Harrowell1,Chen1,Manning1,Xu1}, or geometrically unfavoured motifs (GUMs)~\cite{Ma1}. Unlike crystalline solids whose defects are easily identified from a periodic lattice, no distinguishing structures have been found for the fragile regions of glasses in an apparently disordered background. Without a clear structural indicator for fragile regions in glasses, many studies rely on phenomenological models such as soft glassy rheology (SGR) to understand macroscopic properties of amorphous materials \cite{Sollich1}.

Early studies attempt to connect a particle's propensity to rearrange to its immediate environment. Spaepen proposed a free-volume theory that favors sites with large free volumes for local rearrangements in hard-sphere glasses~\cite{Spaepen1}. For soft sphere systems, which include metallic glasses, Egami and co-workers suggest that regions with extreme local stresses, which correspond to particles with either extremely large or extremely small free volumes, are potential ``defects'' sites~\cite{Egami1,Egami2}. Direct experimental examinations of these scenarios, however, are rare due to the difficulty of measuring local free volumes or atomic stresses in glassy materials~\cite{Huang1}. More recently, machine learning methods are employed to search through multi-dimensional parameter space and use a combination of structural features to identify fragile regions in disordered solids ~\cite{AndreaLiu1, AndreaLiu2}.

Phonon modes, which reflect the collective excitations in glasses, are shown to be a powerful identifier of fragile regions in glasses. Soft spots defined by the low-frequency quasi-localized soft modes are shown to overlap significantly with rearranging regions in glasses~\cite{Chen1, Manning1, Harrowell1}. However, measurements of the local geometric structures of soft spots did not find any distinctive structural characteristic for these regions~\cite{Manning1}. Recent theories and experiments suggest that in glasses, there may exist amorphously correlated structures formed during the glass transition~\cite{Wolynes1,Zamponi1,Tanaka1,Han1,Biroli1,Durian1}, and the dynamics-structure correlation goes beyond single-particle measures~\cite{Berthier1}. Therefore, the search for ``defects'' in glasses needs to probe correlations (or lack thereof) on length scales greater than the first-neighbor shell. Recent simulations by Tong \emph{et al.}~\cite{Xu1}, find strong correlations between the distributions of soft modes in glasses and structural entropy $S_{2}$, raising the possibility that this correlation-based structural parameter may also be able to predict elementary deformations in glasses.

In this letter, we employ video microscopy to directly image local deformations in quasi-2D colloidal glasses consisting of thermo-sensitive microgel particles, and measure the structures of local rearranging regions under cyclic thermal perturbations. Thermal cycling tunes the sizes of soft colloidal particles, and generates uniform compression or dilation when the particles swell or shrink. Similar to mechanical shearing~\cite{Weitz1, Keim1, Candelier1, Regev1, Fiocco1}, rearrangements during thermal cyclings are the results of local stress imbalances. But perturbations from thermal cycling impose no directional bias or boundary effects, and allow simultaneous activation of defects with different orientational preferences. We find that local free volumes are weakly correlated with particle rearrangements; particles with either extremely high or low free volumes are equally likely to rearrange. Local structural entropy $S_{2}$ derived from local pair correlation functions strongly correlates with observed local rearrangements. Rearranging clusters emerge and evolve with regions of high $S_{2}$ values, which are structurally less correlated or more disordered compared to the more stable background, similar to the topological defects in polycrystalline solids. The results presented here are for deeply jammed samples under small periodic perturbations. We have also observed qualitatively the same correlations in samples of different packing fractions and under different perturbation patterns~\cite{suppl.}.

\begin{figure}
    \includegraphics[width=9cm]{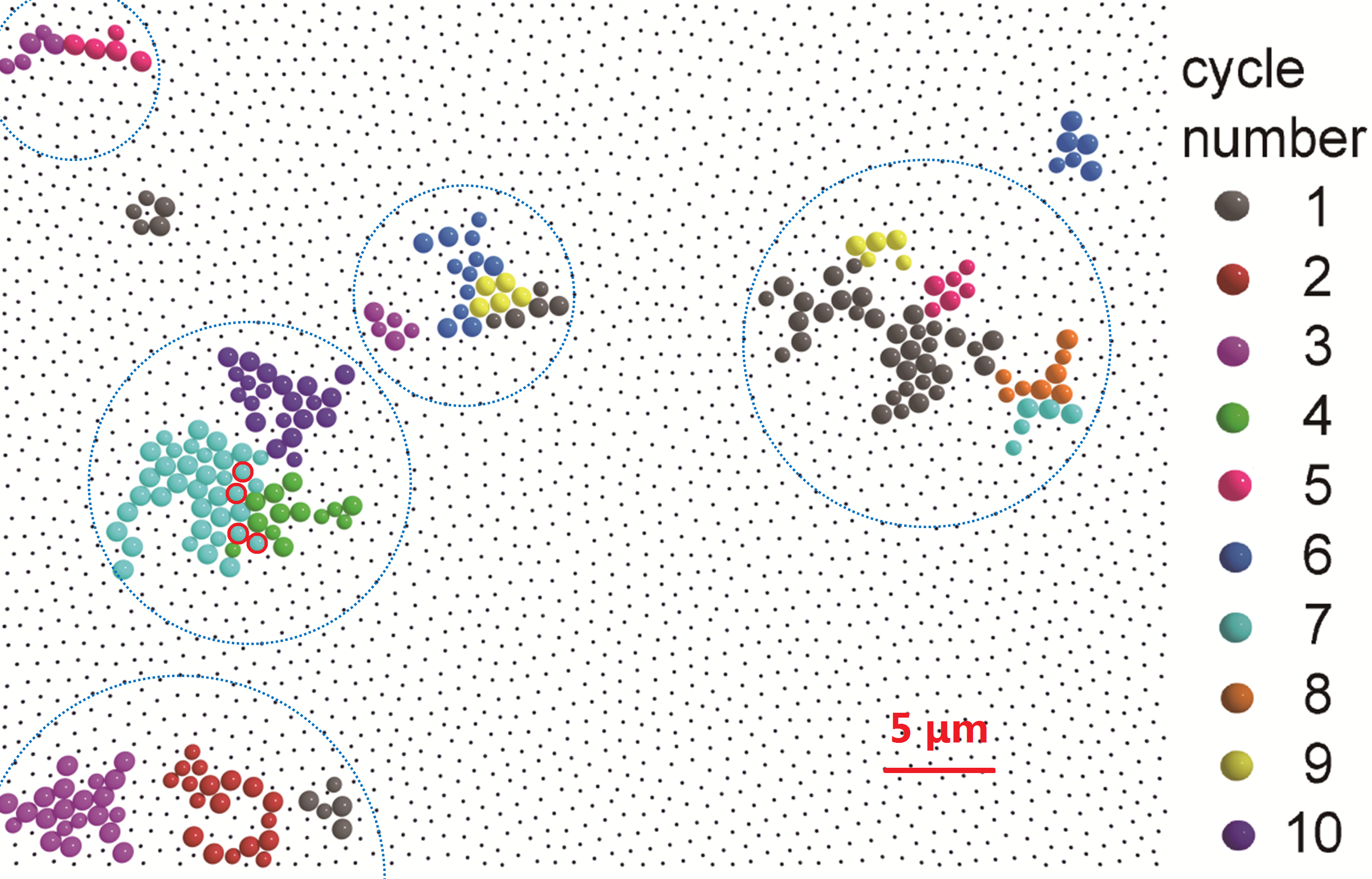}
  \caption{\textcolor[rgb]{0.98,0.00,0.00}{(color online)}  Spatial distribution of rearranging clusters during thermal cycling. Rearranging clusters from different thermal cycle numbers, shown by large spheres in different colors. Small dots are background particles that did not experience neighbor changes during the experiment. Large blue dotted circles indicate regions with repeated rearranging events. Small red circles show the only 4 overlapping rearranging particles during the experiment. }
  \label{fig1}
\end{figure}
The samples are prepared by loading a binary mixture of poly-N-isopropylacrylamide (PNIPAM) particles between two coverslips. The colloidal suspension spreads under capillary forces, and forms a dense monolayer between the glass plates. The samples are then hermetically sealed using optical glue (Norland 63). A binary mixture is used to frustrate crystallization. The diameters of the particles are measured to be $1$ and $1.3~\mu m$ at $22^{\circ}$C by dynamical light scattering; and the number ratio between large and small particles is close to 1. PNIPAM colloidal particles are temperature sensitive whose diameters decrease when temperature increases~\cite{Yunker1,Still1,Chen2,Chen3}. The interactions between PNIPAM particles are generally characterized to be repulsive soft-sphere with a hard core~\cite{Han2}. Rearrangements are induced by locally changing the sample temperature. The colloidal suspension is mixed with a small amount of non-fluorescent dye (Chromatch-Chromatint black 2232 liquid, $0.2\%$ by volume). When illuminated by a mercury lamp, the dye absorbs the incident light and heats up a small area much larger than the field of view, while the rest of the sample remains at ambient temperature~\cite{Han3,Yunker2}. When the mercury lamp is turned off, the heated region rapidly recovers to its original temperature. The temperature increase from optical heating is calibrated to be about $0.2 K$; and the sample reaches new thermal equilibrium in less than 1 second~\cite{suppl.}. For this small temperature change, the samples remain in jammed glassy states at both temperatures, with a brief transient period in between.

The samples are aged on the microscope stage for 200 min. before being periodically heated and cooled for a total of 10 cycles. Each cycle lasts for 30 min., with 15 min. each at the elevated and ambient temperature~\cite{suppl.}. The samples are continuously imaged using standard bright-field microscopy at 20 fps for the duration of the experiment. There are $\sim3700$ particles within the image frame, with a packing fraction of 0.88 at the ambient temperature. The trajectory of each particle is extracted by particle tracking techniques~\cite{Grier1}. We limit our analyses on particles that are at least 3 diameters away from the image boundaries, which leaves $\sim2990$ particles in a reduced field of view. Information of particles outside of this reduced field of view is utilized only for the calculation of parameters for particles within. Most particle rearrangements are observed within 60 seconds of a temperature switch. A video of one local rearranging event can be found in the supplementary materials~\cite{suppl.}

We extract plastically rearranged regions by comparing particle positions at the last minute of each cycle with the positions of the particles 1 minute before the same cycle; at both points the sample is at thermal equilibrium with the ambient temperature. Separation cut-offs defined by the first minima on the particle pair correlation functions between different species are used to determine neighbor changes during thermal cycling~\cite{Harrowell1,Yunker2}. Distributions of particle separations over time are also considered to avoid accidental misidentification~\cite{suppl.}. Particles that changed neighbors during one cycle are grouped into clusters based on nearest-neighbor pairings, and clusters containing fewer than 5 particles are ignored~\cite{Chen1,Manning1}. Most neighbor changes are permanent; only about 2\% of rearranging particles regain their lost neighbors in subsequent thermal cycles.

Fig.~\ref{fig1} shows all rearranging clusters observed during 10 thermal cycles. Clusters from different cycles are shown in different colors. For a single cycle, the spatial distribution of rearranging clusters appears uncorrelated, with cluster separations much larger than typical cluster sizes. However, when the distributions of rearranging clusters from different thermal cycles are compared, it is clear that the clusters are concentrated in certain regions of the sample, as indicated by the large blue dotted circles in Fig.~\ref{fig1}, instead of evenly scattered throughout the sample. Overlapping between neighboring clusters are rare, with only 4 overlapping particles for the duration of the experiment (small red circles in Fig.~\ref{fig1}). The rest of the rearranging particles did not rearrange again during our experiment.

\begin{figure}

    \includegraphics[width=9cm]{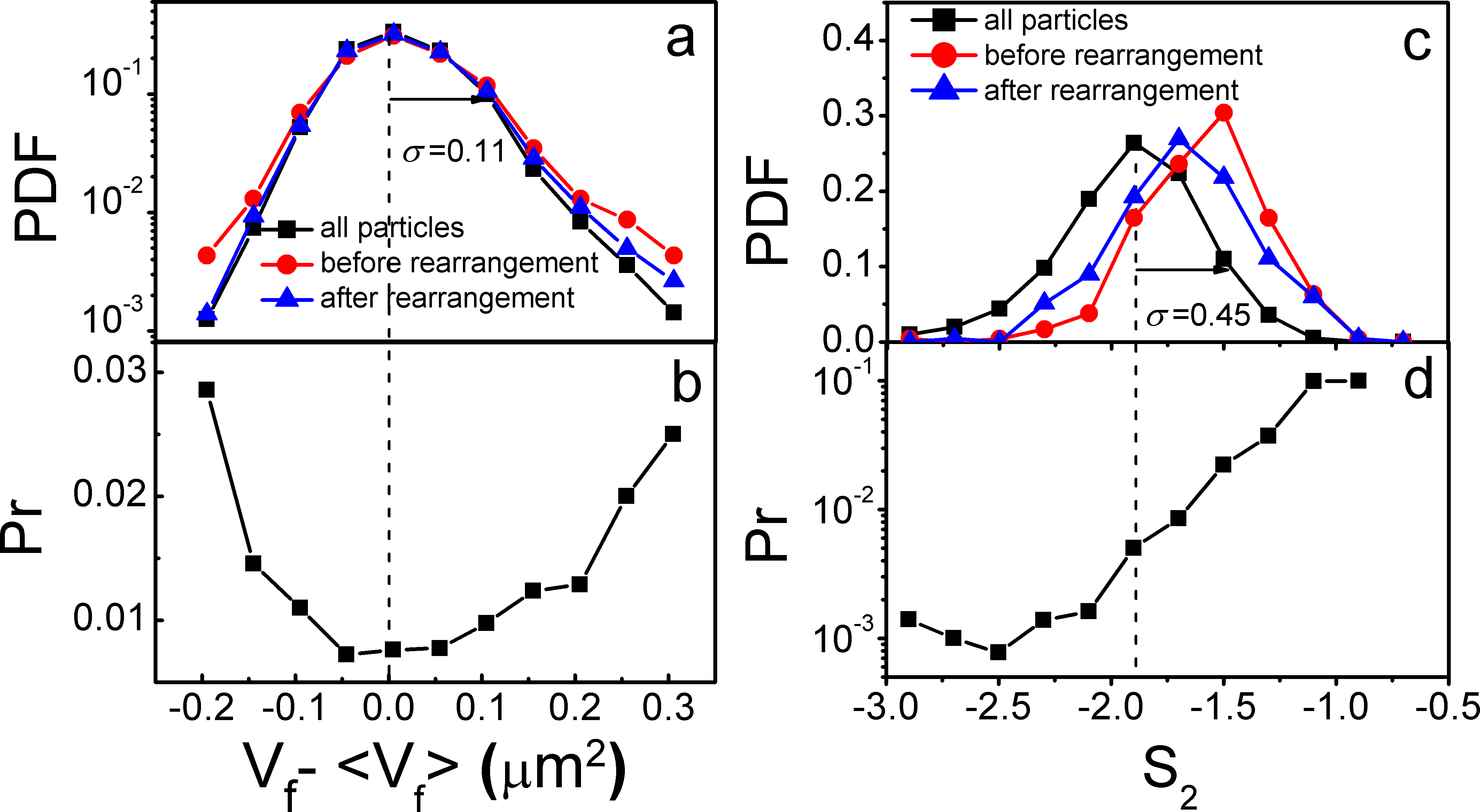}
  \caption{\textcolor[rgb]{0.98,0.00,0.00}{(color online)} Free volumes and local structural entropy in colloidal glasses.  (a) Distribution of $V_{f}$ for all particles (black squares), initial $V_{f}$  for rearranging particles (red circles), and $V_{f}$ of rearranging particles after deformations (blue triangles). (b) Rearranging probability as a function of initial $V_{f}$ (subtracted by the average $V_{f}$ ). (c) $S_{2}$ distribution for all particles (black squares), for rearranging particles before deformations (red circles), and for rearranging particles after deformations (blue triangles). (d) Rearranging probability as a function of initial $S_{2}$. Dashed lines and arrows show the average value and standard deviation of $V_{f}$ and $S_{2}$ for the distributions for all particles. All distributions are obtained from the combined data of 10 cycles, and are individually normalized.}
  \label{fig2}
\end{figure}

We examine the rearranging clusters in search for distinctive structural characteristics from the rest of the colloidal glasses. We first measure the particle free volumes ($V_{f}$). We employ radical Voronoi tessellation to define a polygon for each particle for our binary mixture~\cite{volume1,volume2,volume3}. The radical Voronoi tessellation avoids cutting through large spheres by taking into account the relative particle sizes. The $V_{f}$ is determined by subtracting the cross-section area of a particle from the area of the polygon. The soft-sphere interaction between the colloidal particles allows negative $V_{f}$ for compressed particles. The average $V_{f}$ for particles in a rearranging cluster before and after local rearrangements is effectively the same as that of all the particles in the sample, with slightly larger widths, as plotted in Fig.~\ref{fig2}a. Fig.~\ref{fig2}b plots the measured rearranging probability as a function of initial $V_{f}$ for individual particles. Particles with extremely high or extremely low $V_{f}$ are those most likely to rearrange compared to particles with average $V_{f}$. The curve in Fig.~\ref{fig2}b is roughly symmetric to the system average, suggesting that loose and compact regions are equally prone to rearrangement. This symmetry may explain why previous studies that search for monotonic correlations between local free volumes and rearrangements did not find a significant overlap~\cite{Manning1}.

The rearranging probability for particles with the most extreme free volumes is about 3 times higher than that of the least active particles, which gives free volume relatively weak ability to predict rearranging regions, compared to other known parameters such as soft spots \cite{Chen1, Manning1, Harrowell1}. As a highly local structural parameter, particle free volume focuses only on the contacting neighbors of a particle. Rearrangements in glasses, however, require the cooperation between particles over a longer range~\cite{Harrowell1, Harrowell2}. Thus the structural correlations beyond the first-neighbor shell need to be considered to better identify rearranging regions. Pair correlation function $g(r)$ is a simple measure of structural correlations in glasses. Using local $g(r)$, the structural entropy of a particle \emph{i} can be defined as
$S_{2,i}=-1/2\sum_{\nu}\rho_{\nu}\int
d{\vec{r}}\left\{g^{\mu\nu}_i({\vec{r}}){\rm
    ln}g^{\mu\nu}_i({\vec{r}})-\left[g^{\mu\nu}_i({\vec{r}})-1\right]\right\}$
\cite{Xu1,Tanaka1,Ghosh1}, where $\mu$ and $\nu$ denote the type of particles (large or small), $\rho_{\nu}$ is the number density of $\nu$ particles, and $g^{\mu\nu}_i(\vec{r})$ is the pair correlation function between particle $i$ of type $\mu$ and particles of $\nu$ type. Structural entropy measures the loss of entropy due to positional correlations \cite{Wallace1}; high $S_{2}$ values indicate less correlated local structures and vice versa. In our experiments, the integration is truncated at the $3^{rd}$ neighbor shell to avoid the loss of a large fraction of particles due to boundary effect. We verify, for particles near the center of the field of view, that the $S_{2}$ obtained from 3 shells of neighbor are highly correlated to $S_{2}$ obtained from 5 neighbor shell integrations, with correlation coefficients about 0.9. Beyond the $5^{th}$ neighbor shell, local $g(r)$ becomes very close to 1, thus contributing only negligibly to measured $S_{2}$. For each particle, time-average is performed for local $g(r)$ before integration to remove short-time fluctuations, although in principle, this averaging is not critical to the measurements of $S_{2}$~\cite{Tanaka2}. We limit the averaging time to the last 60 seconds (1200 frames) of each thermal cycle, within the $\beta$ relaxation time of the sample, to avoid direct coupling to diffusive dynamics~\cite{Han1,Tanaka2}.

\begin{figure}[b]

  {\includegraphics[width=9cm]{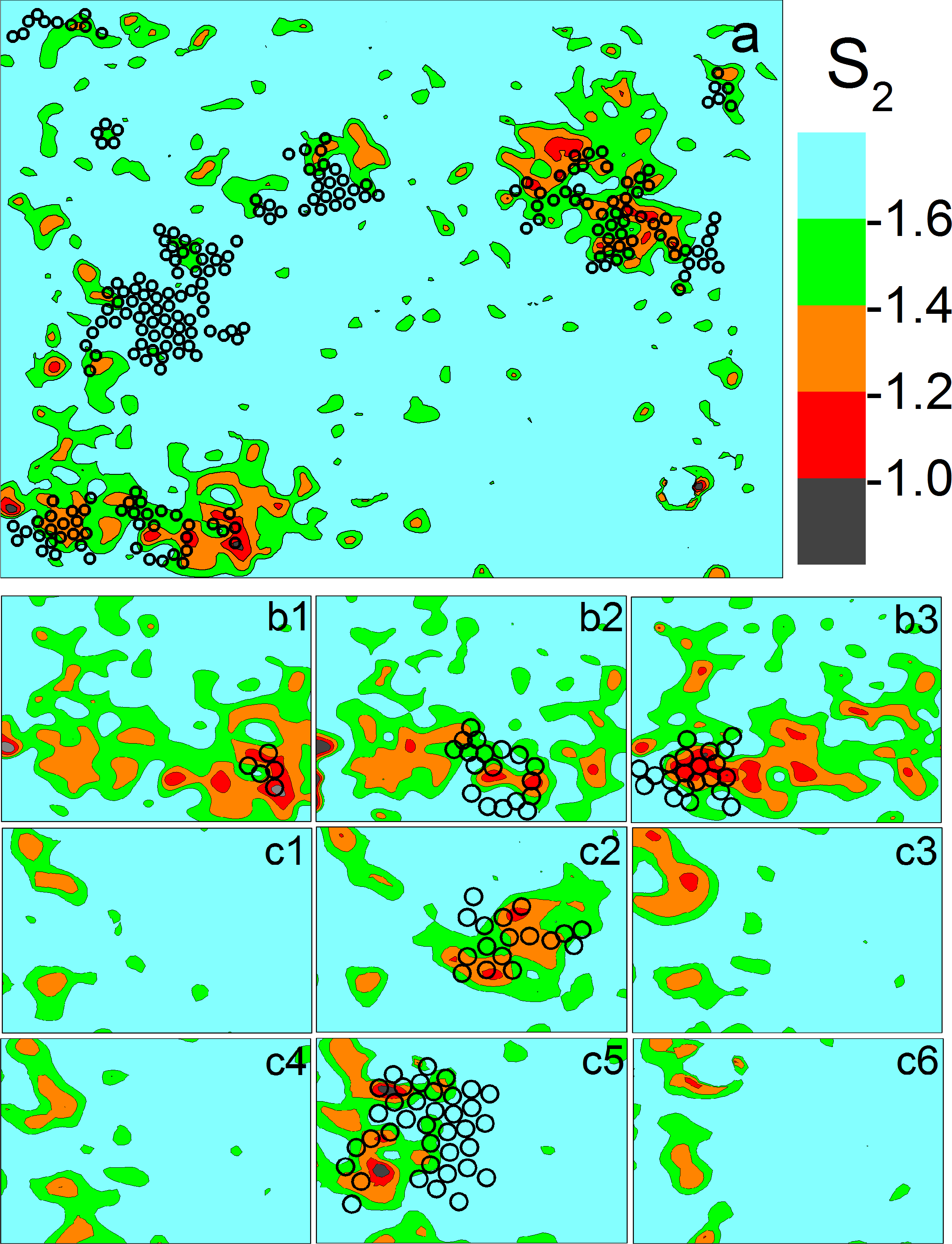}}
  \caption{\textcolor[rgb]{0.98,0.00,0.00}{(color online)} Spatial distribution of particle $S_{2}$ and rearranging clusters. (a) Colored contour plot: $S_{2}$ distribution \emph{before} thermal cycling, and rearranging clusters observed in the following 10 cycles (black circles). (b1-b3) Contour plots: local $S_{2}$ distributions before the $1^{st}$,$2^{nd}$, and $3^{rd}$ thermal cycle in a sub-region in the sample; black circles: rearranging clusters from $1^{st}$,$2^{nd}$ and $3^{rd}$ in the same region. (c1-c6) Contour plots: local $S_{2}$ distributions from the $3^{rd}$ to $8^{th}$ thermal cycle in another sub-region in the sample; black circles: rearranging clusters from the $4^{th}$ cycle (c2), and the $7^{th}$ cycle (c5).}
  \label{fig3}
\end{figure}

The structural entropy of rearranging particles are qualitatively different from those of the general population in the colloidal glass. Fig.~\ref{fig2}c plots the distribution of $S_{2}$ for all the particles in the colloidal glass (black squares) and those for rearranging particles (red circles). Compared to $V_{f}$ distributions, the distribution of $S_{2}$ for rearranging particles peaks at a significantly higher value than the distribution from all particles, indicating less correlated structures in rearranging clusters. After rearrangements, the $S_{2}$ values of particles in rearranging clusters are decreased, but remain elevated compared to the system average, as shown by blue triangles in Fig.~\ref{fig2}c.  Fig.~\ref{fig2}d plots the rearranging probability as a function of initial $S_{2}$ for individual particles. The probability to rearrange increases almost monotonically with $S_{2}$, with two orders of magnitude difference between the most stable particles and most unstable particles, revealing intrinsically different responses from local structures in colloidal glasses.~Qualitatively the same correlations as shown in Fig. 2 have been observed in samples of different packing fractions and under different perturbation patterns~\cite{suppl.}.

Similar to soft modes, the structural entropy for individual particles can be employed to predict regions prone to rearrangements. The colored contour plot in Fig.~\ref{fig3}a shows the initial distribution of $S_{2}$ in the colloidal glass \emph{before} thermal cycling, black circles overplotted on the contour plot are the rearranging clusters observed in the following 10 cycles. Remarkable overlap is observed between regions with high $S_{2}$ values and rearranging clusters, which demonstrates that local $S_{2}$ is able to predict rearranging regions long before they actually occur in colloidal glasses under thermal cycling. This correlation is enhanced when the system is closer to the relaxation events, as shown in the video of the evolution of the $S_{2}$ distributions and the rearranging clusters from each thermal cycle in the supplementary materials~\cite{suppl.}.

The overlap between the $S_{2}$ distribution before thermal cycling and rearranging clusters in later cycles suggests that the distribution of high $S_{2}$ regions evolves slowly under thermal cycling, which can be clearly seen in the video in the supplementary materials~\cite{suppl.}. An example of the stability of local distribution of $S_{2}$ is shown in Fig.~\ref{fig3}(b1-b3) (down-left corner of the video in the supplementary materials~\cite{suppl.}). During the 3 consecutive local rearrangements, the area with high $S_{2}$ values only changes slightly. On the other hand, high $S_{2}$ regions can be created through local fluctuations. Some rearranging clusters that have small overlaps with high $S_{2}$ regions in Fig.~\ref{fig3}a are found to significantly overlap with high $S_{2}$ spots that emerge during the thermal cycling. An example is shown in Fig.~\ref{fig3}(c1-c6) (mid-left region of the video in the supplementary materials~\cite{suppl.}). Before the 4th and 7th cycles, two high $S_{2}$ spots are developed with no preceding rearranging clusters in its immediate neighborhood. The emergences of these two spots are followed by two rearranging events in the same region, after which local $S_{2}$ distribution falls back to the background.

We quantitatively evaluate the correlations between high $S_{2}$ regions and local rearrangements during thermal cycling. High $S_{2}$ clusters are identified to the particle level by applying a cut-off to the distribution of local structural entropy. Averaged correlation between high $S_{2}$ clusters and rearranging clusters is measured to be 0.34 for a cutoff value of -1.5~\cite{suppl.}, comparable to the correlations between soft spots and local rearrangements in colloidal experiments~\cite{Chen1}. The highest single-cycle correlation is measured to be 0.69~\cite{suppl.}. We also measure the direct correlations between non-affine displacement coefficient $D_{min}^2$ and $S_{2}$ for all particles, regardless of neighbor changes. $D_{min}^2$ measures the particle level non-affine strain, and is defined as the minimum of $D^2(t_{1},t_{2})=\sum\limits_{n}\sum\limits_{i}[r^i_{n,t_{2}}-r^i_{0,t_{2}}-\sum\limits_{j}(\delta_{ij}+\varepsilon_{ij})\times(r^j_{n,t_{1}}-r^j_{0,t_{1}})]^2$
, where $r^i_{n,t}$ is the $i$-th ($x$ or $y$) component of the position of the $n$-th particle at time t (before or after a single cycle) and index $n$ runs over the particles within the interaction range of the reference particle $n=0$. The $\delta_{ij}+\varepsilon_{ij}$ that minimizes $D^2$ are calculated based on $r^i_{n,t}$~\cite{Langer1}. Direct correlation between $D_{min}^2$ and particle $S_{2}$ for all particles at each cycle yields an averaged correlation coefficient of 0.30~\cite{suppl.}, suggesting that particles with higher $S_{2}$ values tend to experience larger strains under external loading. The noise levels for the above two correlation coefficients are both on the order of $\sim1/\sqrt{N}$, where $N$ is the number of particles in the field of view.

\begin{figure}

  {\includegraphics[width=8cm]{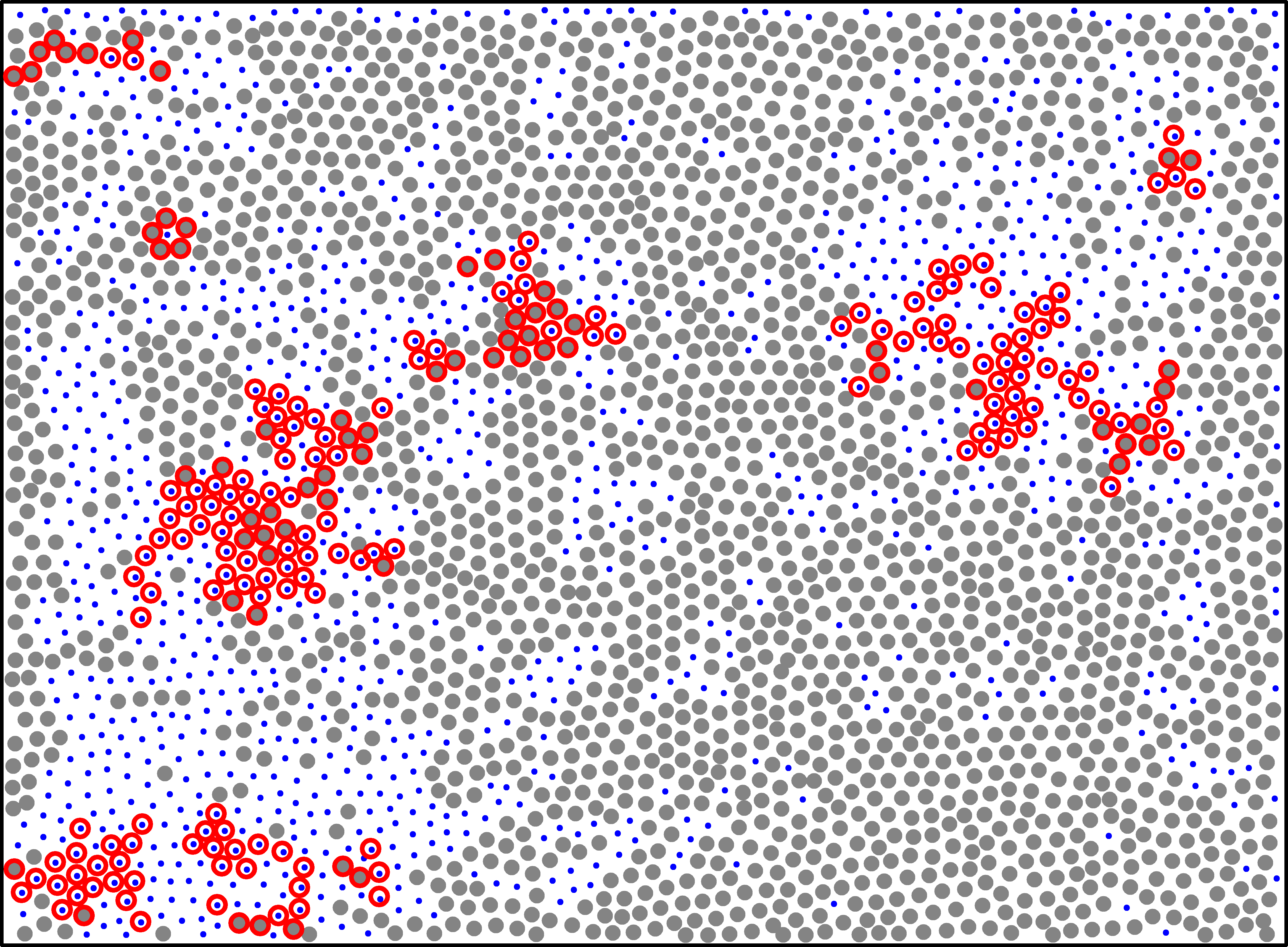}}
  \caption{\textcolor[rgb]{0.98,0.00,0.00}{(color online)} Stable regions in colloidal glasses. Spatial distributions of particles with $S_{2}$ values below -1.6 during the experiment (large filled gray circles), particles with $S_{2}$ values exceeding -1.6 for at least one cycle (small blue dots), and rearranging clusters (red empty circles).}
  \label{fig4}
\end{figure}

On the other hand, structurally stable regions associated with low $S_{2}$ values in the colloidal glasses show strong resistance to the migration of rearranging clusters. Fig.~\ref{fig4} plots the particles whose $S_{2}$ remain below -1.6 (the $90^{th}$ percentile in the $S_{2}$) for the duration of the experiment (large gray circles). These grey particles form a large percolating network that only overlaps with rearranging clusters on the boundaries with high $S_{2}$ areas (small blue dots). The structural stability of these correlated domains may be related to the amorphously ordered structures predicted by random first-order transitions theories~\cite{Wolynes1,Zamponi1}, and the dynamical slowdown during glass transitions~\cite{Tanaka1,Han1, T1}, which are fundamental challenges in the physics of glasses. And our results suggest $S_{2}$ may be employed to search for such structures.

We can now make the analogy between structural defects in crystals to high $S_{2}$ regions in colloidal glasses. In polycrystalline solids, structural correlation is strong between atoms belonging to the same domain, and weak for atoms on defect sites. Similarly, the high $S_{2}$ spots in colloidal glasses are structurally less correlated compare to the percolating low $S_{2}$ domains. Thus high $S_{2}$ regions can be viewed as equivalent ``structural defects'' in glasses. This analogy is supported by the  mechanical susceptibility of these regions and their creation and migration dynamics observed in the experiments.

To summarize, we measure the structures of local rearranging clusters in colloidal glasses under thermal cycling. The local structural entropy of the rearranging regions is distinctively higher than that of the more stable background in glasses. Equivalent defects, that share the essential characteristics of structural defects in crystals, can be defined by the high $S_{2}$ values. As a structural parameter, $S_{2}$ provides a sensible measure of the apparently disordered structures of glasses. The strong connections between the structural entropy to physical observables such as system dynamics \cite{Tanaka1,Ghosh1,Mittal1} and local relaxations suggest that structural correlations may play a critical role in the physics of glasses. Investigation of more complex structural correlations such as multi-body correlations~\cite{Tanaka2} in glasses may reveal deeper connections between structures and the macroscopic properties of glasses.

We thank Yiwu Zong, Baoan Sun, Chris H. Rycroft for instructive discussions. This work was supported by the MOST 973 Program (No. 2015CB856800). K. C. also acknowledges the support from the NSFC (No. 11474327).

\end{document}